\documentclass{article}
\usepackage{spconf,amsmath,amssymb,graphicx}
\usepackage[T1]{fontenc}
\usepackage[utf8]{inputenc}
\usepackage{mathptmx}
\usepackage{booktabs}
\usepackage{multirow}
\usepackage{cite}
\usepackage[hidelinks]{hyperref}
\usepackage{balance}
\usepackage{tikz}
\usetikzlibrary{arrows.meta, positioning, fit}

\title{Extracting Voice Styles from Frozen TTS Models via Gradient-Based Inverse Optimization}
\name{Gyeongmin Kim}
\address{Department of Computer Science, Hanyang University, Seoul, South Korea\\
kdrkdrkdr@hanyang.ac.kr}
\begin{document}
\ninept
\maketitle
%
%
\begin{abstract}
Some text-to-speech systems ship a synthesis model and preset style vectors but withhold the reference encoder that turns a recording into a style vector, so a user cannot obtain a style for a new voice. We recover that vector without the encoder by inverting the released pipeline with gradient descent: all weights stay frozen and only the style vector is optimized, against time-pooled WavLM statistics of one recording of the target. The objective discards the time axis, so no transcript is needed. Two constraints from the release make the result a drop-in style: every row of the timbre style stays on the unit sphere where the presets lie, and the small duration style, which a time-pooled loss cannot see, is fitted to the recording's speaking rate. On SupertonicTTS, over 147 speakers and 100 held-out sentences each, ECAPA-TDNN similarity rises from 0.129 to 0.419, every recovered style starts from a preset and ends closer to its target than that preset was, and the pooled word error rate stays below that of the presets. A verifier at its equal-error point accepts 52\% of the recovered voices as the target speaker, against 1\% of the presets.
\end{abstract}


\section{Introduction}
\label{sec:intro}

Neural text-to-speech (TTS) models such as SupertonicTTS~\cite{supertonic2025} synthesize speech far faster than real time with a fraction of the parameters of earlier systems. Many condition generation on a \textit{style vector} that encodes speaker identity and vocal character, so one model can voice any speaker for which such a vector exists. The vector for a new speaker normally comes from a \textit{reference encoder}, called a speaker encoder in some systems. Some released systems ship the synthesis model and a handful of preset styles but keep this encoder private. The synthesizer still takes a style vector. Only the map from audio to it is missing. Producing the vector for a given recording without the encoder is model inversion~\cite{fredrikson2015inversion}: the function is fixed and public, and the unknown is one of its inputs.

We recover the missing input by optimization: given the released pipeline and one target recording, we search for the style vector whose synthesized output matches the target voice, with nothing trained and nothing in the release modified. Three difficulties follow from the missing encoder. Gradients have to pass through an inference graph never meant for training. The objective has to measure speaker identity while the generated and the target utterance say different words. And with no encoder there is no description of where valid styles lie, nor of how far the loss should be driven.

Fitting a conditioning vector while the generator is held fixed is a known form of speaker adaptation~\cite{taigman2018voiceloop, arik2018cloning, chen2019sample}, but the published methods adapt a model their authors trained, and they supervise the fit by reconstructing the target's own transcribed utterance frame by frame. Here the generator is someone else's inference-only release and the target is a bare recording. We therefore discard the time axis and match per-dimension statistics of an early self-supervised learning (SSL) layer, which are content-independent up to a measurable floor (Section~\ref{ssec:stop}).

Our contributions are:
\begin{enumerate}
\item A procedure that recovers a withheld style vector from a released synthesizer. Nothing is trained, and no transcript, forced aligner, speaker-verification model, enrolled population, or reference encoder is needed.
\item What the release itself supplies. The presets reveal the geometry of the style space, which the recovered vector is kept on, and the loss floor at which to stop. They also give the duration style a starting point for matching the speaking rate.
\item An evaluation on 147 speakers from two corpora, on 100 held-out sentences per speaker. Code and demo:\newline \mbox{\href{https://github.com/kdrkdrkdr/supertonic.embed}{\nolinkurl{github.com/kdrkdrkdr/supertonic.embed}}}.
\end{enumerate}

\begin{figure*}[t]
\centering
\begin{tikzpicture}[
    node distance=3mm,
    box/.style={draw, rounded corners=2pt, minimum height=6.5mm,
                minimum width=19mm, inner xsep=1mm, inner ysep=1pt,
                align=center, fill=white},
    inp/.style={box, fill=black!10},
    opt/.style={box, fill=black!28, very thick},
    arr/.style={-{Stealth[length=2mm]}, thick},
    darr/.style={-{Stealth[length=2mm]}, thick, dashed},
  ]
  \node[opt] (sttl) {$\mathbf{s}_{\text{ttl}}$};
  \node[box, right=5mm of sttl] (te) {Text\\encoder};
  \node[box, right=of te] (ve) {Vector\\estimator};
  \node[box, right=of ve] (voc) {Vocoder};
  \node[inp, right=of voc] (gen) {Generated\\audio};
  \node[inp, right=5mm of gen] (wavlm) {WavLM\\layer 4};
  \node[box, right=of wavlm] (loss) {$\mathcal{L}(\boldsymbol{\mu},\boldsymbol{\sigma})$};
  \node[inp, above=2mm of wavlm] (tgt) {Target audio};
  \node[box, above=7.5mm of voc] (proj) {Unit-sphere\\projection};
  \node[opt, below=7mm of sttl] (sdp) {$\mathbf{s}_{\text{dp}}$};
  \node[box, right=5mm of sdp] (dp) {Duration\\predictor};
  \node[box, below=7mm of gen] (rate) {Speaking-rate\\match};

  \node[draw, dashed, rounded corners=3pt, inner sep=2.5mm,
        fit=(te)(ve)(voc), label={[yshift=0.5mm]above:Frozen TTS pipeline}] (pipe) {};

  \draw[arr] (sttl) -- (te);
  \draw[arr] (te) -- (ve);
  \draw[arr] (ve) -- (voc);
  \draw[arr] (voc) -- (gen);
  \draw[arr] (gen) -- (wavlm);
  \draw[arr] (tgt) -- (wavlm);
  \draw[arr] (wavlm) -- (loss);
  \draw[darr] (loss.north) |- (proj.east);
  \draw[darr] (proj.west) -| node[pos=0.25, above] {$\nabla_{\mathbf{s}_{\text{ttl}}}\mathcal{L}$} (sttl.north);
  \draw[arr] (sdp) -- (dp);
  \draw[arr] (dp.north) -- ++(0,2.5mm) -| node[pos=0.5, below] {latent length} (ve.south);
  \draw[arr] (gen.south) -- (rate.north);
  \draw[darr] (rate.west) -- node[below] {fit $\rho$} (dp.east);
\end{tikzpicture}
\caption{The timbre style $\mathbf{s}_{\text{ttl}}$ (dark) receives the gradient of the time-pooled WavLM loss and is projected back onto the unit sphere after every step. The duration style $\mathbf{s}_{\text{dp}}$ sets the latent length and is fitted so that the speaking rate of the synthesis matches the target. All modules of the TTS pipeline and of WavLM remain frozen; generated and target audio may carry different text.}
\label{fig:overview}
\end{figure*}
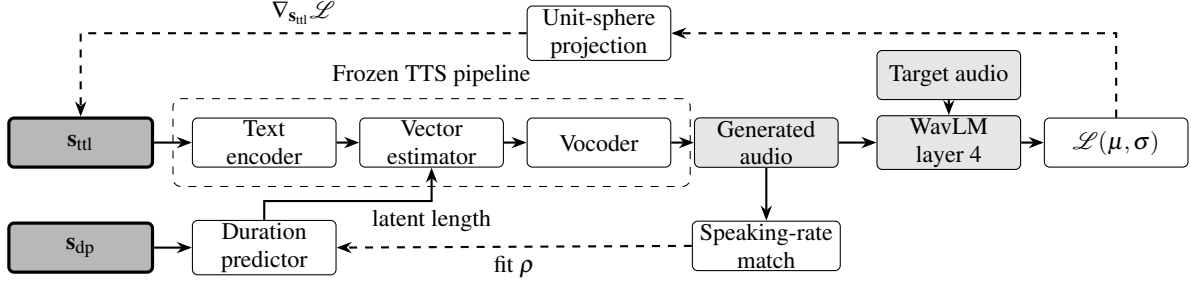

\section{Related Work}
\label{sec:related}

\textbf{Cloning and adaptation.} Zero-shot systems obtain a voice from a speaker encoder~\cite{jia2018transfer} or by prompting on codec tokens~\cite{wang2023valle}. Closer to our setting, Voice\-Loop~\cite{taigman2018voiceloop}, the embedding-only variant of Arik et al.~\cite{arik2018cloning}, and SEA-EMB~\cite{chen2019sample} keep the generator fixed and fit a conditioning vector for one speaker. They differ from our case in two respects. The generator is one the authors trained, so its weights and corpus are theirs to use. And the objective reconstructs the target's own transcribed utterance with a per-frame loss, which, as the VoiceLoop paper notes, ``requires an exact temporal alignment of the input and the output sequence.'' SEA-EMB also needs linguistic features and $f_0$ from the adaptation data. None of this is available for a bare recording or accepted by an inference-only release.

\textbf{Optimizing a conditioning vector toward other goals.} Other work optimizes inside a frozen speech generator without touching its weights: a style code driven until an automatic speech recognition (ASR) system returns a chosen transcript~\cite{jin2024style}, or a speaker embedding driven toward endless generation~\cite{gao2024ttslow}. Neither objective refers to a target voice. Closest to us, Marras et al.~\cite{marras2023dictionary} optimize the conditioning embedding of a frozen cloner to maximize speaker-embedding similarity to a proxy population, an approach they call ``untargeted dictionary attacks.'' The vector is made to pass as many speakers at once rather than to reproduce one, and the objective needs a speaker-verification encoder, which our method never uses.

\textbf{Reaching a voice without a speaker encoder.} kNN-VC~\cite{baas2023knnvc} and kNN-TTS~\cite{hajal2024ssltts} reach a chosen voice from untranscribed audio and public self-supervised features, but by retrieval: the target is kept as a database of frames, not reduced to a vector, and nothing is optimized toward it. Aldeneh et al.~\cite{aldeneh2024pooled} use the time-pooled mean and standard deviation of a frozen self-supervised layer, as our loss does, but as a similarity score between two real recordings.

\textbf{Latent optimization.} GAN inversion~\cite{abdal2019image2stylegan} recovers latent codes through a frozen generator. Where alignment is meaningless, the image domain replaces the pixel loss with a pooled statistic: Image2StyleGAN++~\cite{abdal2020i2s2} has a mode that optimizes the latent against a style loss alone, with its pixel losses switched off. Audio style transfer~\cite{grinstein2018audiostyle} matches such statistics too, but optimizes the output signal rather than an input to a fixed generator. A layer-wise probe of WavLM-Large places peak speaker-identification accuracy at layer~4~\cite{chiu2025probing}.

\section{Method}
\label{sec:method}

The deployment pattern we target is a TTS release that publishes the synthesis model and preset style vectors while keeping the reference encoder private. Few systems fit it. Most either ship the speaker encoder with the weights (CosyVoice~\cite{du2024cosyvoice}), prompt directly on reference audio without an explicit style vector (F5-TTS~\cite{chen2024f5tts}), or condition on discrete codec tokens (VALL-E~\cite{wang2023valle}). We study SupertonicTTS, which fits the pattern and is small enough to differentiate through end to end (Fig.~\ref{fig:overview}).

\subsection{Problem formulation}

Let $G$ be a frozen TTS pipeline mapping text $t$ and a style $\mathbf{s}$ to a waveform. For a target recording $\mathbf{y}^*$ we solve
\begin{equation}
    \mathbf{s}^* = \arg\min_{\mathbf{s}} \; \mathcal{L}\big(G(t, \mathbf{s}),\ \mathbf{y}^*\big),
\end{equation}
where $t$ is text of our choosing, unrelated to what $\mathbf{y}^*$ says. SupertonicTTS exposes a timbre style $\mathbf{s}_{\text{ttl}} \in \mathbb{R}^{50 \times 256}$, which enters the acoustic modules as cross-attention keys and values, and a duration style $\mathbf{s}_{\text{dp}} \in \mathbb{R}^{8 \times 16}$, which enters only the duration predictor and sets how long the utterance runs. We recover both: the 12{,}800 parameters of $\mathbf{s}_{\text{ttl}}$ by the gradient of Eq.~\ref{eq:loss}, the 128 of $\mathbf{s}_{\text{dp}}$ by fitting the speaking rate of the recording afterwards (Section~\ref{ssec:dp}).

\subsection{Gradient flow through a deployed graph}

The released pipeline consists of four ONNX modules built for inference. We simplify each graph with onnxslim, lower the opset from~19 to~17, and remove trailing empty inputs from Clip operators, after which onnx2torch yields a differentiable PyTorch module. All converted parameters and all WavLM parameters are kept frozen.

\subsection{A content-independent objective}
\label{ssec:obj}

The generated and the target utterance contain different words, so the objective has to ignore content. Let $\mathbf{H} = [\mathbf{h}_1, \dots, \mathbf{h}_T] \in \mathbb{R}^{T \times D}$ denote the WavLM-Large layer-4 hidden states, $D{=}1024$. We collapse the time axis into per-dimension statistics,
\begin{equation}
    \boldsymbol{\mu} = \tfrac{1}{T}\textstyle\sum_{\tau} \mathbf{h}_\tau, \qquad
    \boldsymbol{\sigma}^{2} = \tfrac{1}{T-1}\textstyle\sum_{\tau} (\mathbf{h}_\tau - \boldsymbol{\mu})^{2},
\end{equation}
the square taken elementwise, and compare the synthesized utterance $\hat{\mathbf{y}} = G(t, \mathbf{s})$ with the target $\mathbf{y}^*$ by the mean squared error
\begin{equation}
    \mathcal{L} = \tfrac{1}{D}\big\|\boldsymbol{\mu}_{\hat{y}} - \boldsymbol{\mu}_{y^*}\big\|_2^2
                + \tfrac{1}{D}\big\|\boldsymbol{\sigma}_{\hat{y}} - \boldsymbol{\sigma}_{y^*}\big\|_2^2 .
\label{eq:loss}
\end{equation}
This is the statistics pooling used in x-vector speaker embeddings~\cite{snyder2018xvector}. Even taken directly from frozen SSL states, without a trained head, these statistics separate the speakers of a clean corpus at an equal-error rate of a few percent~\cite{aldeneh2024pooled}, which is presumably why the objective transfers to verification metrics it never saw. We use layer~4 because Chiu et al.~\cite{chiu2025probing} report the highest speaker-identification accuracy for this model there. Their indexing counts the CNN encoder output as layer~0, as ours does.

\subsection{Staying on the manifold, and where to stop}
\label{ssec:stop}

With the encoder unavailable, the region of valid styles is documented only by the presets. Every one of the 500 token rows across the ten released $\mathbf{s}_{\text{ttl}}$ vectors has unit L2 norm, to within $2{\times}10^{-7}$, so each row lies on the unit sphere $S^{255} \subset \mathbb{R}^{256}$ and the withheld encoder evidently writes onto the product of 50 such spheres. Free gradient descent has no reason to stay on that surface, so we project every row back to unit norm after each optimizer step. The rows of $\mathbf{s}_{\text{dp}}$ are unit vectors in the presets as well, and the fit in Section~\ref{ssec:dp} keeps them so.

The presets also fix the stopping point. Two different sentences synthesized from the same preset leave a residual loss that is not due to the speaker. Over the ten presets and the five prompts of Section~\ref{ssec:opt}, these 100 same-preset pairs give losses between $0.148$ and $0.413$, mean $0.242$. Inside that band the objective can no longer tell a different sentence from a different speaker. We stop as soon as the loss reaches $0.24$, the mean of the band.

\subsection{Optimization details}
\label{ssec:opt}

Optimization starts from the nearest preset, the one of the ten released presets whose output on the first prompt is closest to the target under Eq.~\ref{eq:loss}. Five English prompts chosen for phoneme coverage rotate from step to step, so the style vector cannot overfit one phoneme set. We use Adam with learning rate $2{\times}10^{-4}$, gradient clipping at max-norm~$1.0$, and ReduceLROnPlateau with patience~200 and factor~$0.5$. Each prompt has its own noise latent, drawn once from a fixed seed and reused at every step and for every speaker. Its length is set once from the predicted duration of one preset, so during this stage only the timbre differs between two speakers. Since each style vector influences only its own output, several speakers are optimized in one forward pass by summing their losses, and a speaker that reaches the threshold leaves the batch immediately for a waiting one. With a batch of~16 on one RTX~3090, the timbre stage takes about half a minute per speaker.

\subsection{Recovering the duration style}
\label{ssec:dp}

A time-pooled loss discards the time axis, which is where speaking rate lives, so $\mathbf{s}_{\text{dp}}$ cannot be recovered by the same gradient. Its entire audible effect is a single number, the length the duration predictor assigns to an utterance, and we fit this number to the recording from audio alone. The speaking rate of a clip is measured as the number of peaks of its smoothed RMS envelope per second of trimmed speech, a crude syllable-nucleus count that needs no transcript. The same detector runs on the target and on the synthesis, so its undercount cancels in the ratio.

Starting from the preset's $\mathbf{s}_{\text{dp}}$, we look for the multiplier $\rho$ on its predicted durations at which the synthesis speaks at the target's rate. Each iteration has four steps. It inverts the frozen duration predictor by gradient descent so that it outputs the current target durations, with a small penalty toward the starting vector and the eight rows projected to unit norm. It synthesizes the five prompts at the latent length this $\mathbf{s}_{\text{dp}}$ commands, as inference does. It measures their rate. And it takes a secant step in log-duration toward the target. $\rho$ itself is not clamped, but one iteration may change the duration by at most a factor of two, and up to eight iterations are allowed, a budget no speaker exhausted. The iteration stops when the rates agree within 3\%, when two consecutive moves fail to improve the match, or when the length has moved by about a fifth without the rate following it. The fit takes about a minute per speaker.

\section{Experiments}
\label{sec:exp}

\subsection{Setup}

We evaluate the public SupertonicTTS~2 release~\cite{supertonic2release} (internal version \texttt{v1.6.0}), whose architecture is described in~\cite{supertonic2025}. It ships 10 preset styles and no reference encoder.

Targets come from two corpora with different recording conditions: 110 speakers from VCTK~\cite{vctk} (studio, 8 to 12\,s of reference audio, gender and accent metadata) and 44 from the Common Voice subset of Seed-TTS Eval~\cite{seedtts2024} (crowdsourced, 3 to 16\,s). Because the objective pools over time, a reference that is mostly silence describes the room rather than a voice. We keep a reference only if its measured speaking rate over the whole file is at least 2.2 nuclei per second and the mean RMS of its loudest fifth of frames is at least 22\,dB above that of its quietest fifth. The rule looks at the reference alone, never at what the extraction produced. It removes 7 of the 110 VCTK references (six below the rate threshold, two below the noise threshold, one below both), and all numbers below are over the remaining 147 speakers. The kept references all lie clear of both thresholds, the lowest at 2.44 nuclei per second and 23.4\,dB, and the same rule leaves all 44 Seed-TTS references in place.

For every speaker we synthesize 100 Harvard sentences (IEEE 1969, lists 1 to 10), none of them among the five optimization prompts, from the extracted style and from the preset it started from, so the two are compared on the same sentences. The recording's own words are never used. Speaker similarity is the cosine similarity between the speaker embeddings of the reference and of each generated clip, averaged over the 100 sentences. We compute it under ECAPA-TDNN~\cite{desplanques2020ecapa} (SIM\textsubscript{E}) and a ResNet (SIM\textsubscript{R}), both VoxCeleb models from SpeechBrain, and under WavLM-base-plus-sv (SIM\textsubscript{W}), a WavLM-family verifier of the kind many TTS papers report similarity with. The first two are trained independently of our objective, and the third shares the WavLM family with it. Intelligibility is the word error rate (WER) under Whisper-large-v3, pooled over all words of a system. We then read the transcripts and removed differences that are not errors, using one fixed list applied to every system: digits against number words, seven compound spellings, three spelling variants, apostrophes, and three homophone pairs.

For scale, pairs of different speakers, which a verifier calls impostor pairs, score $0.100$ ECAPA similarity on average over 6000 pairs of these references, with a 95th percentile of $0.302$ and a maximum of $0.554$. A second recording of the same VCTK speaker scores a median of $0.680$. The nearest preset starts at the impostor floor, and since the impostor tail is long, the comparison that matters is the paired change per speaker.

\subsection{Main results}
\label{ssec:results}

\begin{table}[t]
\centering
\caption{147 speakers, 100 held-out sentences each; the Preset row is the preset each speaker started from, on the same sentences. WER is adjudicated (raw pooled over all speakers: 7.24\% preset, 7.00\% extracted). Bold marks the better of the two.}
\label{tab:results}
\small
\setlength{\tabcolsep}{3.5pt}
\begin{tabular}{@{}llcccc@{}}
\toprule
& & \textbf{SIM\textsubscript{E}}$\uparrow$ & \textbf{SIM\textsubscript{R}}$\uparrow$ & \textbf{SIM\textsubscript{W}}$\uparrow$ & \textbf{WER}$\downarrow$ \\
\midrule
\multirow{2}{*}{All (147)} & Preset & 0.129 & 0.095 & 0.703 & 6.00\% \\
 & \textbf{Extracted} & \textbf{0.419} & \textbf{0.409} & \textbf{0.837} & \textbf{5.71\%} \\
\midrule
\multirow{2}{*}{VCTK (103)} & Preset & 0.128 & 0.092 & 0.691 & 5.73\% \\
 & \textbf{Extracted} & \textbf{0.416} & \textbf{0.406} & \textbf{0.833} & \textbf{4.37\%} \\
\midrule
\multirow{2}{*}{Seed-TTS (44)} & Preset & 0.130 & 0.102 & 0.731 & 6.65\% \\
 & \textbf{Extracted} & \textbf{0.427} & \textbf{0.415} & \textbf{0.847} & 8.86\% \\
\bottomrule
\end{tabular}
\end{table}

\begin{figure*}[t]
\centering
\includegraphics[width=\textwidth]{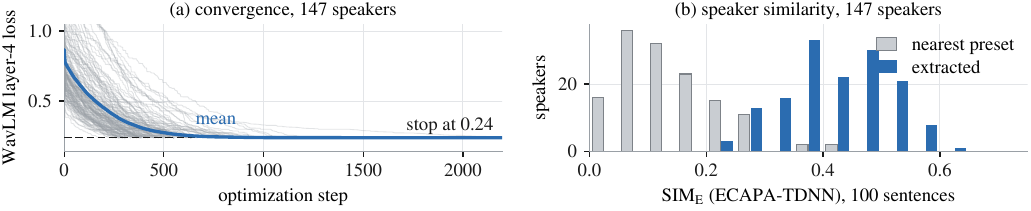}
\caption{(a) Best-so-far loss for all 147 speakers (light: individual speakers, dark: their mean); every speaker reaches 0.24, after a median of 460 steps (range 190 to 2197). (b) ECAPA similarity over the 100 sentences before and after extraction.}
\label{fig:results}
\end{figure*}

Table~\ref{tab:results} compares the extracted styles with the preset each started from, the nearest of the ten released presets under Eq.~\ref{eq:loss}. Speaker similarity rises from the impostor floor to SIM\textsubscript{E} 0.419 and SIM\textsubscript{R} 0.409 on sentences the optimization never saw. Bootstrap 95\% confidence intervals over speakers are $[0.405, 0.434]$ and $[0.395, 0.423]$. On the five prompts the style was fitted to, SIM\textsubscript{E} is 0.475, and on five further held-out prompts of similar length 0.467, so new text costs about a hundredth.

The rise holds for every speaker: all 147 recovered styles end closer to their target than their starting preset under SIM\textsubscript{E} and SIM\textsubscript{R}, and 141 of 147 under SIM\textsubscript{W}. The objective sees none of these three metrics. A Wilcoxon signed-rank test gives $p < 10^{-25}$ for both independent metrics, with paired effect sizes $d_z$ of 3.44 and 3.46, and the two verification models rank speakers similarly (Spearman $\rho = 0.857$).

Intelligibility is not affected overall. Pooled over about 114{,}000 words, the extracted voices are transcribed at 5.71\% word error rate against 6.00\% for the presets they started from, and the per-speaker median falls from 3.88\% to 2.84\%. The 6\% floor belongs to the frozen model. SupertonicTTS occasionally drops or repeats words, and in our measurement the same sentence from the same preset can move by eighteen points of WER between noise seeds. This is why we report word error pooled over 100 sentences. Extraction raises the word error rate only on the Seed-TTS subset, where the fitted duration styles move furthest from the presets, and a tempo pushed toward the edge of the model's range costs words.

The recovered duration style does what it was fitted for. On the five prompts used by the fit, the synthesized speaking rate deviates from the recording by 9.4\% with the preset's $\mathbf{s}_{\text{dp}}$ and by 2.0\% with the recovered one; on five held-out prompts the error falls from 11.9\% to 8.1\%. The multipliers range from 0.61 to 1.30, and 26 speakers move by more than a fifth. Similarity is unchanged (SIM\textsubscript{E} 0.477 before the fit, 0.475 after), as expected, since $\mathbf{s}_{\text{dp}}$ reaches the audio only through its length. Only about half of the reduction in rate error carries over to new text, because nuclei per second also depend on how dense the sentence is.

Predicted naturalness (UTMOS~\cite{saeki2022utmos}) over the same sentences is 4.03 for the extracted voices against 4.27 for the presets, that is 94\% of the preset score, and above the 3.92 of the real references under the same predictor. The median speaker loses 0.11 ($d_z = 0.68$, against 3.44 for the change in similarity); VCTK speakers keep 98\% of the preset score, Seed-TTS speakers 86\%, so the drop falls where the word errors do. UTMOS was trained on earlier systems, so only the paired comparison should be read.

\subsection{Would a verifier accept these voices?}
\label{ssec:verify}

\begin{table}[t]
\centering
\caption{Acceptance by an ECAPA verifier tuned to a given false-accept rate. Genuine: a second real recording of the same speaker (103 VCTK speakers); impostor: 6000 pairs over all 147 references; extracted: each speaker's mean over the 100 sentences.}
\label{tab:far}
\small
\begin{tabular}{@{}lcccc@{}}
\toprule
Operating point & Thresh. & Genuine & \textbf{Extracted} & Preset \\
\midrule
EER (1.0\%) & 0.410 & 99.0\% & \textbf{51.7\%} & 1.4\% \\
FAR = 5\% & 0.302 & 100.0\% & \textbf{89.1\%} & 2.7\% \\
FAR = 1\% & 0.417 & 98.1\% & \textbf{47.6\%} & 0.0\% \\
FAR = 0.1\% & 0.527 & 95.1\% & 11.6\% & 0.0\% \\
\bottomrule
\end{tabular}
\end{table}

A cosine similarity is difficult to interpret as a risk, so Table~\ref{tab:far} restates the result as a verification decision: with the verifier tuned to a given false-accept rate (FAR) on genuine and impostor trials, how often does it accept an extracted voice as the target speaker? The verifier separates real speakers well, with an equal-error rate (EER) of 1.0\% at a threshold of 0.410. At that operating point it accepts 51.7\% of the extracted voices and 1.4\% of the presets they were built from; at a 5\% false-accept rate, 89.1\% against 2.7\%.

VCTK's gender labels also allow the initialization to be checked. The nearest preset matches the labeled gender for 98 of 103 speakers, and all five errors are one male preset (M1) chosen for a female speaker. The descent recovers from the wrong start: those five also end well above the impostor floor.

\section{Discussion and Limitations}
\label{sec:limit}

We adopt the layer that Chiu et al.~\cite{chiu2025probing} report as best for speaker identity. Neighboring early layers would plausibly work too, but each would need its own content floor and stopping point. The rate detector cancels its own bias but not the effect of the sentence, so a rate matched on five prompts transfers only partly to other sentences. Normalizing by the syllable count of the text would narrow that gap, at the price of needing a transcript on the target side. We measure acceptance by a speaker verifier, not by a spoofing countermeasure. Synthetic speech from this model may well be detectable even when the verifier accepts it, and our result only bounds how much identity information the conditioning space carries. Finally, the results come from a single TTS model, naturalness is a predicted score rather than a listening test, and the weakest results fall on the crowdsourced Seed-TTS subset.

\section{Conclusion}
\label{sec:conclusion}

We have recovered a withheld style vector from a released synthesizer by optimization alone. A time-pooled early-layer WavLM statistic makes the comparison content-independent up to a measured floor, so one untranscribed recording is enough, and the released presets provide the rest: the surface the style vector has to stay on, the floor at which to stop, and a starting point for the speaking rate. Across 147 speakers and 100 held-out sentences each, the method raises ECAPA similarity from 0.129 to 0.419, moves every speaker from its preset closer to its target, does not cost intelligibility overall, and takes about ninety seconds of GPU time per voice. A verifier at its equal-error point accepts 52\% of the recovered voices as the target speaker, against 1\% for the presets. Withholding the encoder removed a convenience but not the capability, and we think a partial release should be judged by what its released components already determine.

\section{Compliance with Ethical Standards}

\textbf{Use of a released model outside its intended interface.} SupertonicTTS ships synthesis weights and preset voices and withholds the reference encoder; we reach the conditioning space without it. We used only public artifacts, circumvented no access control, and modified no released file, but operating a release outside its intended interface still carries obligations beyond its license.

\balance
\textbf{Dual use.} The measurement is directly usable to imitate a real person: what we report is how convincingly a recovered vector reproduces a specific voice, so a high number is also what makes misuse feasible. We report it because the same number tells a provider, or a user relying on one, how much of a voice a partial release already determines.

\textbf{Data, approval and funding.} All evaluation uses consented, openly licensed corpora: VCTK (CC BY 4.0) and the Common Voice subset of Seed-TTS Eval (CC0). The study was retrospective, on openly available speech data. No new recordings were collected and ethical approval was not required. No identifiable individual is targeted and no public figure appears. The only extracted style vectors we release belong to the five example speakers whose CC0 Common Voice recordings ship with the code; they are published with the demo page, and both the code and the page tell users to obtain explicit consent from anyone whose voice they clone. This work received no external funding, and the author declares no conflict of interest.

\textbf{What we release, and mitigations.} We publish the procedure and the code that produced the numbers; the technique they build on has been public since 2018~\cite{taigman2018voiceloop, arik2018cloning, chen2019sample}. The conditioning space is reachable because it is continuous, differentiable, and shipped with working examples. Perturbing or quantizing the released presets, watermarking synthesizer output, or serving the acoustic modules behind an interface that hides gradients would each raise the cost of what we did. We measured none of these.

\bibliographystyle{IEEEbib}
\bibliography{refs}

\end{document}